\documentclass[graybox,natbib,nosecnum]{svmult}
\bibpunct{(}{)}{;}{a}{}{,} 

\pdfoutput=1   

\usepackage{mathptmx}       
\usepackage{helvet}         
\usepackage{courier}        
\usepackage{type1cm}        

\usepackage{makeidx}         
\usepackage{graphicx}        
\usepackage{multicol}        
\usepackage[bottom]{footmisc}
\usepackage[normalem]{ulem}	
\usepackage{hyperref}  

\usepackage{soul}   

\makeindex             

\begin{document}

\title*{Occurrence Rates from Direct Imaging Surveys}
\author{Brendan P. Bowler and Eric L. Nielsen}
\institute{Brendan P. Bowler \at Department of Astronomy, The University of Texas at Austin, Austin, TX 78712, USA, 
\email{bpbowler@astro.as.utexas.edu}
\and Eric L. Nielsen \at Kavli Institute for Particle Astrophysics and Cosmology, Stanford University, Stanford, CA 94305, USA, 
\email{enielsen@stanford.edu}}
%
%
\maketitle

\abstract{
The occurrence rate of young giant planets from direct imaging surveys 
is a fundamental tracer of the efficiency with which planets form and migrate
at wide orbital distances.
These measurements have progressively converged to a value of 
about 1\% for the most massive planets ($\approx$5--13~$M_\mathrm{Jup}$)
averaged over all stellar masses at separations spanning a few tens to a few hundreds of AU.
The subtler statistical properties of this population are beginning to emerge with ever-increasing 
sample sizes:
there is tentative evidence that planets on wide orbits are more frequent around
stars that possess debris disks; 
brown dwarf companions exist at comparable (or perhaps slightly higher)
rates as their counterparts in the planetary-mass regime; and 
the substellar companion mass function appears to be smooth and may extend down 
to the opacity limit for fragmentation.
Within a few years, the conclusion of second-generation direct imaging surveys will enable more definitive interpretations 
with the ultimate goal of identifying the dominant origin of this population 
and uncovering its relationship to planets at smaller separations.
}

\section{Introduction}

Direct imaging is the foremost method to study giant planets at wide orbital distances beyond about 10 AU
and complements  
radial velocity, transit, microlensing, and astrometric discovery techniques
which probe smaller separations closer to their host stars.
High-contrast imaging from the ground makes use of adaptive optics systems that largely operate at
near-infrared wavelengths, making this method most sensitive 
to thermal emission from massive, warm giant planets.
As a result, direct imaging surveys predominantly focus on the closest and youngest stars before
planets have cooled to faint luminosities and low temperatures.
This makes target samples for imaging surveys unusual compared to other planet detection methods, which
predominantly focus on old (several Gyr) field stars with lower activity and jitter levels.

\begin{figure}
\hskip -.7in
\includegraphics[scale=.65]{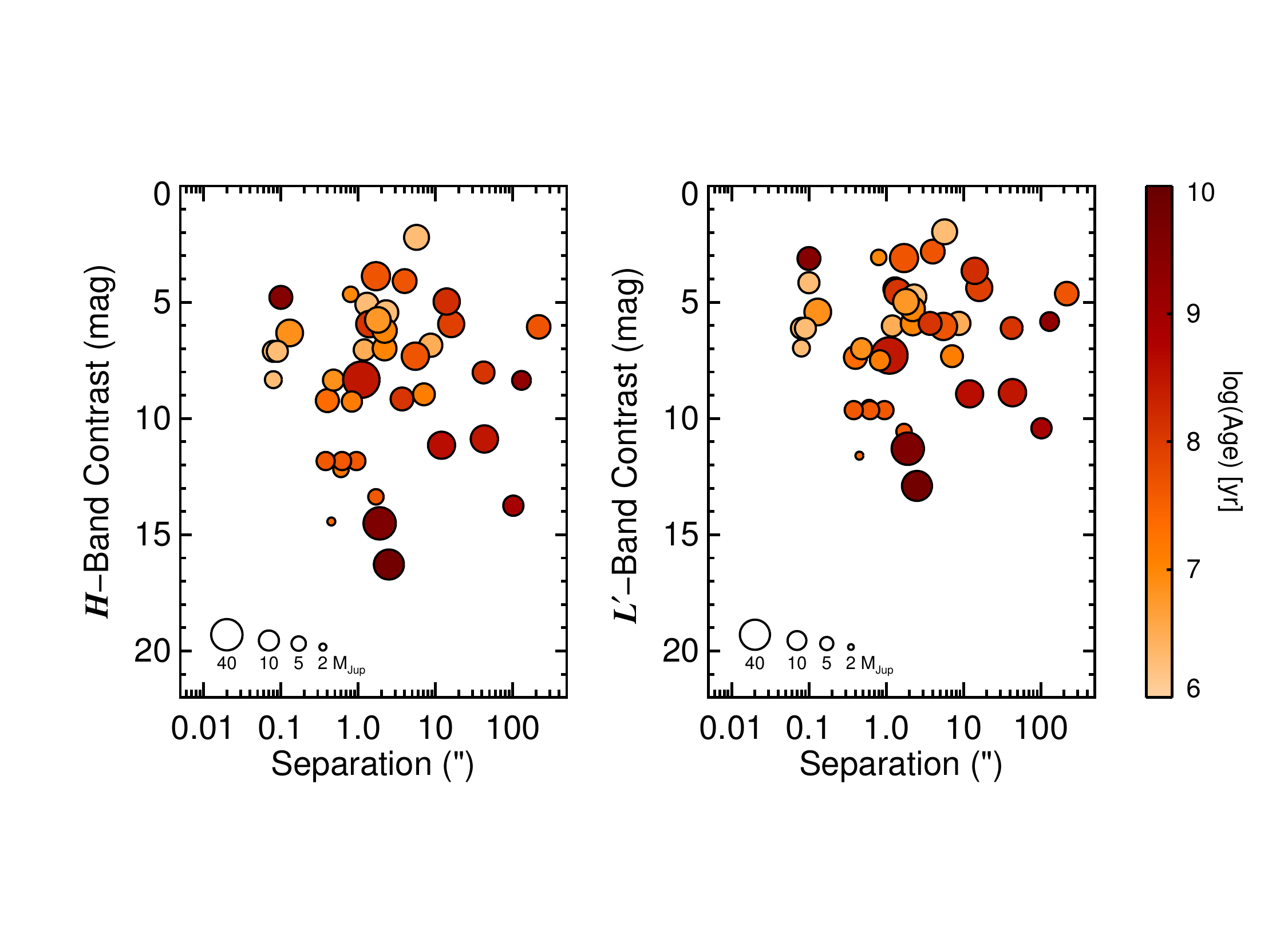}
\vskip -.6in
\caption{Imaged companions with masses near or below the deuterium-burning limit.
Objects are from \citet{Bowler:2016jk} and are supplemented with 
recently-discovered companions from \citet{Bowler:2017hq} and \citet{Chauvin:2017hl}.  
Photometry represent actual measurements or values taken from the hot-start evolutionary models 
of \citet{Baraffe:2015fwa}
based on the inferred companion mass and age, here represented by the size and color of the symbol.}
\label{fig:absmagsep}       
\end{figure}

In addition to individual discoveries, high-contrast imaging surveys deliver statistical constraints on the occurrence rates and 
demographics of giant planets at large orbital distances.
The frequency of giant planets and their mass-period distributions provide valuable information about the efficiency of planet
formation and migration to large separations, and have been used to both guide and test giant planet formation routes. 
Nearly two dozen young planets have now been imaged with 
inferred masses as low as 2 $M_\mathrm{Jup}$ and separations spanning a large
dynamic range of $\approx$10--10$^4$ AU (Figure~\ref{fig:absmagsep}).
Several formation routes can explain the origin of gas giants at these unexpectedly wide separations:
core and pebble accretion (\citealt{Lambrechts:2012gr}), 
disk instability (e.g., \citealt{Durisen:2007wg}; \citealt{Kratter:2016dw}),
turbulent fragmentation (\citealt{Bate:2003cv}), and planet-planet scattering (\citealt{Veras:2009br}).  
In principle these pathways should 
imprint unique signatures on the resulting occurrence rates, mass-period distributions, 
and three-dimensional orbital architecture of exoplanets, although  these are challenging to discern in practice 
due to the low incidence of widely-separated planets, a diversity of theoretical predictions, the
the potential for subsequent migration to occur, and the difficulty of constraining orbital
elements for ultra-long period planets (e.g., \citealt{Blunt:2017eta}).
One of the emerging goals for direct imaging is to untangle the dominant formation route(s) of this population,
which is best accomplished in a statistical fashion with expansive high-contrast imaging surveys.

The largest-scale surveys exploiting extreme adaptive optics systems like 
the Gemini Planet Imager (GPI) and the Spectro-Polarimetric High-contrast Exoplanet REsearch (SPHERE) 
are currently underway.
These second-generation instruments utilize integral field spectrographs
for speckle suppression through spectral differential imaging and achieve
unprecedented on-sky contrasts at small ($<$1$''$) angular separations using coronagraphs.
These results will eventually be merged with 
first-generation surveys to provide statistics of giant planets for samples exceeding one
thousand young stars.
In the future, the James Webb Space Telescope, the Extremely Large Telescopes,
and space-based telescopes with coronagraphs working in the optical like WFIRST 
(and perhaps HabEx or LUVOIR) will expand 
this parameter space to lower masses, closer separations, and older ages.

Calculating occurrence rates for direct imaging surveys is a unique challenge.  Surveys yield few detections, 
which means the underlying planetary mass-semimajor axis distributions are poorly constrained.  This differs from
other methods where detections are plentiful and the intrinsic functional form of the planet size, mass, and period
distributions can be precisely  measured.  Instead, several assumptions must be made--- or details ignored---
about the underlying demographics and the way in which planets may scale with stellar mass.

Here we provide an overview of the methods and assumptions most widely adopted to calculate
occurrence rates from direct imaging surveys, as well as a summary of observational 
results from the largest adaptive optics imaging
surveys and meta-analyses carried out to date in the brown dwarf ($\approx$13--75 $M_\mathrm{Jup}$) and planetary ($<$13  $M_\mathrm{Jup}$)
mass regimes.

\section{Calculating Occurrence Rates}

High-contrast adaptive optics imaging is sensitive to angular scales spanning an inner working angle,
often set by the size of a coronagraph, and an outer working angle limited by the field of view of
the detector.
The primary products of an observation is astrometry and relative photometry of point sources as well as
a contrast curve in $\Delta$mag (flux ratio) which
defines the sensitivity of an observation at some statistical threshold.
Point sources may be bound companions or (more often the case) background stars, which are distinguished
via their relative motion or colors.
Transforming high-contrast imaging observations to  
giant planet frequencies  
relies on a series of assumptions about the brightness of 
planets in particular filters
as well as the underlying shape of the planet mass, separation, 
and eccentricity distributions.
Here we provide an outline of the typical end-to-end procedure carried out by most surveys: 

$\bullet$  \textit{A survey is designed and first epoch observations are acquired.}

Direct imaging surveys have been continually increasing in size and sensitivity over the past 15 years.
Target samples have evolved from several dozen stars to several hundred in size,
although this is ultimately limited by the number of nearby young stars.
Similarly, representative limiting contrasts have improved from $\approx$10~mag at 0.5$''$ (\citealt{Biller:2007ht})
to $>$14 mag with second-generation extreme AO systems (e.g., \citealt{Samland:2017hl}).  
Recent surveys generally boast a strategic advantage over previous work by targeting newly-identified 
nearby young stars, taking advantage of novel speckle suppression methods, or probing closer 
inner working angles with innovative coronagraph designs.
The most successful surveys with the highest impacts are those that aim to answer specific 
science questions with clear and measurable goals.  
Reconnaissance work is an important component to designing an efficient survey, 
for example by vetting close visual binaries using adaptive optics on small telescopes, 
or by reassessing stellar ages to prioritize the youngest stars.  
First-epoch observations are acquired for all targets with typical integration times of about an hour each.

$\bullet$  \textit{Speckle subtraction is carried out, point sources are identified, and limiting contrasts are calculated.}

Several methods for PSF subtraction have been developed over the past decade.  Their aim is largely the same: to remove the static diffraction pattern and residual speckle noise from the host star while minimizing self-subtraction from real point sources.  The most widely-used algorithms are variations of the Locally-Optimized Combination of Images (LOCI; \citealt{Lafreniere:2007bg})--- a least-squares approach which aims to represent  science frames as linear combinations of reference images--- and Karhunen-Lo\`{e}ve Image Projection (KLIP; \citealt{Soummer:2012ig})--- a technique that utilizes principal component analysis to identify and remove quasi-static PSF structure.  For integral field units and dual-channel imagers, PSF subtraction takes advantage of the wavelength-dependency of speckles, which are magnified (and better-corrected) at longer wavelengths.  An incomplete list of advances, modifications, and upgrades to these algorithms can be found in \citet{Marois:2010hs}; \citet{Pueyo:2012ft}, \citet{Brandt:2013in}, \citet{Currie:2014hn}, \citet{Marois:2014ep}; \citet{Wahhaj:2015jz}, \citealt{Cantalloube:2015km}, \citet{Wang:2016gl}, \citet{Pueyo:2016hl}, 
and \citet{Ruffio:2017ev}.

Residual speckle noise can resemble planets, so automated point source identification is ideal to minimize biases and develop quantitative, measurable, and reproducible results based on detection threshold metrics like signal-to-noise ratio or reduced $\chi^2$ value (\citealt{Wahhaj:2013fq}).  Reporting contrast curves for null results and detections alike are critical for deriving statistical constraints.  Approaches involving injection-recovery of point sources as a function of separation, contrast, and azimuthal angle offer a robust means of deriving sensitivity limits.
Special care must be taken when calculating confidence levels at small separations where the number of independent resolution elements diminishes (\citealt{Mawet:2014ga}).
Reporting the fractional field of view coverage (from 100\% sensitivity down to 0\%) is also recommended and ideally should be incorporated in the survey statistical analysis.
More recently, \citet{JensenClem:2018jk} propose a more formal strategy in the form of \emph{performance maps}, which are based on signal detection theory and incorporate the true positive fraction of planets, the false positive fraction, and detection thresholds (see also \citealt{Ruffio:2017ev}).

$\bullet$  \textit{Second-epoch observations are taken to determine the nature of point sources.}

Second-epoch observations are usually required to distinguish between background stars and 
comoving, gravitationally-bound companions, which may also exhibit some orbital motion.
This assumes that background stars have zero (or negligible) proper motion compared to
typical astrometric uncertainties (several milli-arcseconds).  However, 
Nielsen et al. (2017) showed that the close comoving point source near HD 131399 (\citealt{Wagner:2016gs}) may instead be an unfortunate example of a background star with a similar proper motion.
Even with second epochs, targets with low proper motions can appear to be comoving and 
statistical false alarm probabilities should regularly be reported.
On the other hand, if colors or spectral information are acquired during first-epoch observations then false alarm probabilities can be calculated using stellar space densities of ultracool dwarfs to statistically validate late-type companions (e.g., \citealt{Macintosh:2015ewa}).

Given limitations of finite telescope time and the fact that new candidate planets can be revealed in follow-up imaging means that surveys often end without having determined the nature of some candidate planets.
Since these untested point sources will be bound planets or background stars, it is imperative that these sources be documented, reported, and treated in a consistent manner when calculating occurrence rates.
A new Direct Imaging Virtual Archive (DIVA) database has been created to assemble 
processed images, published contrast curves, and candidate companions (see \citealt{Vigan:2017kb} for details).

$\bullet$  \textit{Sensitivity limits in planet mass and separation are determined using evolutionary models.}

Monte Carlo simulations of synthetic planets on random orbits offer a straightforward 
approach to transform contrast curves into sensitivity maps in mass and separation--- that is,
the probability that a planet with a given semi-major axis and mass would have been detected in a given
observation or set of observations.  
Uncertainties in a star's age and distance can naturally be folded into this type of analysis.  
Another advantage of the Monte Carlo method is that if a target has been 
observed on multiple occasions spanning more than one epoch, 
these contrast curves can readily be used to 
jointly constrain the properties of potential companions.  
Circular orbits are often assumed, but the underlying orbital eccentricities of the simulated 
planets can alter the final results.
This requires the use of evolutionary models 
to transform a planet's mass and age into an apparent brightness in a particular filter.
Commonly used ``hot-start'' models which ignore radiative losses during planet assembly
include Cond models (\citealt{Baraffe:2003bj}) and BT-Settl models (\citealt{Baraffe:2015fwa}).
``Cold-start'' prescriptions will produce comparatively  pessimistic limiting masses as they
take into account dissipative accretion luminosity during the epoch of planet formation (e.g., \citealt{Fortney:2008ez})

$\bullet$  \textit{Occurrence rates are calculated assuming an underlying planetary mass-period distribution.}

A universal goal of direct imaging surveys is to constrain both the underlying shape and overall
occurrence rate of giant planets in order to evaluate the demographics of giant planets,
measure the efficiency of planet formation, and test for correlations with stellar parameters.
Over two dozen companions near and below the deuterium-burning boundary have been
discovered with direct imaging (e.g., \citealt{Bowler:2016jk}),
but only a handful of these were found in published surveys so the intrinsic underlying distribution 
of giant planets at large separations is not well known.  

What is clear is that planetary-mass objects reside in a wide range of contexts: they have been found 
orbiting high-mass stars (\citealt{Marois:2008ei}; \citealt{Lagrange:2010fs}; \citealt{Macintosh:2015ewa}),
low-mass stars (\citealt{Bowler:2017hq}), brown dwarfs (\citealt{Chauvin:2005gg}; \citealt{Todorov:2010cn}), 
nested within debris disks (e.g., \citealt{Rameau:2013ds}),
multiple systems (\citealt{Delorme:2013bo}), 
free-floating without a host (e.g., \citealt{Liu:2013gya}), 
and even as binary pairs themselves (\citealt{Best:2017bra}).
These companions exist over a wide range of parameter space spanning 
five orders of magnitude in orbital distance
and merge with the population of higher-mass brown dwarf companions, making their
mass, separation, and eccentricity distributions difficult to unambiguously characterize.

With only limited knowledge of the planet distribution function, and without compelling 
physical motivation for an alternative model at wide separations (although see \citealt{Meyer:2017wh}), 
a smooth double power law in mass and semimajor axis is commonly used:

\begin{equation}
\frac{d^2N}{dM da} \propto M^{\alpha} a^{\beta}
\end{equation}

\noindent An inner and outer cutoff in semi-major axis can be applied to ensure the distribution is  constrained
over the interval in which imaging data exists (typically between $\approx$5--500~AU). Close binaries are generally
avoided or removed from these samples.  The planet distribution function can also be modified to take into account
multiple planet systems like HR 8799 (e.g., \citealt{Wahhaj:2013iq}).
The absence of similar systems
strongly implies that the probability of a star hosting a planet depends on whether another planet exists in the
system; that is, they are not independent events (see, e.g., \citealt{Brandt:2014cw}) .

Information about the eccentricity distribution is slowly trickling in via orbit monitoring
of imaged planets.  
Except for the putative companion to Fomalhaut, which has $e$ = 0.8$\pm$0.1 
(\citealt{Kalas:2013hpa}; \citealt{Beust:2014dj}), all imaged planets with orbit constraints 
are inconsistent with high eccentricities: 
HR 8799 bcde and $\beta$ Pic b are all on low-eccentricity orbits with $e$$<$$\sim$0.3 
(e.g., \citealt{Macintosh:2014js}; \citealt{Nielsen:2014js}; \citealt{Wang:2016gl}; 
\citealt{Konopacky:2016gv}; \citealt{Wertz:2017bi}),
51~Eri~b is only marginally constrained at $e$$<$0.7 (\citealt{DeRosa:2015jla}),
and HD 95086 b has $e$ $<$ 0.44 (\citealt{Rameau:2016dx}).
With the exception of $\beta$~Pic b, these planets have been observed for significantly 
less than an orbital period, so in most cases their eccentricities are poorly constrained.

Detections and nondetections can be treated as Bernoulli trials and occurrence rates can be calculated with 
binomial statistics by taking into account the planet sensitivity maps (e.g., \citealt{Lafreniere:2007cv}; \citealt{Nielsen:2010jt}; \citealt{Bowler:2015ja}).
Alternatively, planet frequencies can be calculated by integrating under the fitted planet distribution function
(e.g., \citealt{Brandt:2014cw}; \citealt{Clanton:2016ft}).
In general, a decision must be made about how to treat point sources that have not been confirmed or refuted
as planets.  One possibility is to truncate the contrast curve for those particular targets at some threshold above
the candidate companion to exclude that information from the survey (\citealt{Nielsen:2013jy}; 
\citealt{Bowler:2015ja}; \citealt{Meshkat:2017jka}).

\citet{Bonavita:2013ia} have created a publicly-available package to carry out 
statistical analysis of direct imaging surveys--- Quick Multi-purpose Exoplanet Simulation System (QMESS)---
which offers a computationally fast, flexible, grid-based statistical tool for measuring occurrence rates
(see also the Monte Carlo-based package from \citealt{Bonavita:2012dc}).
\citet{Brandt:2014cw} offer an alternative, analytic framework to transform a set of
survey observations into statistical constraints on the companion frequency and mass function.

\section{Occurrence Rate of Giant Planets on Wide Orbits}

\begin{figure}
\vskip -.6in
\hskip -.7in
\includegraphics[scale=.65]{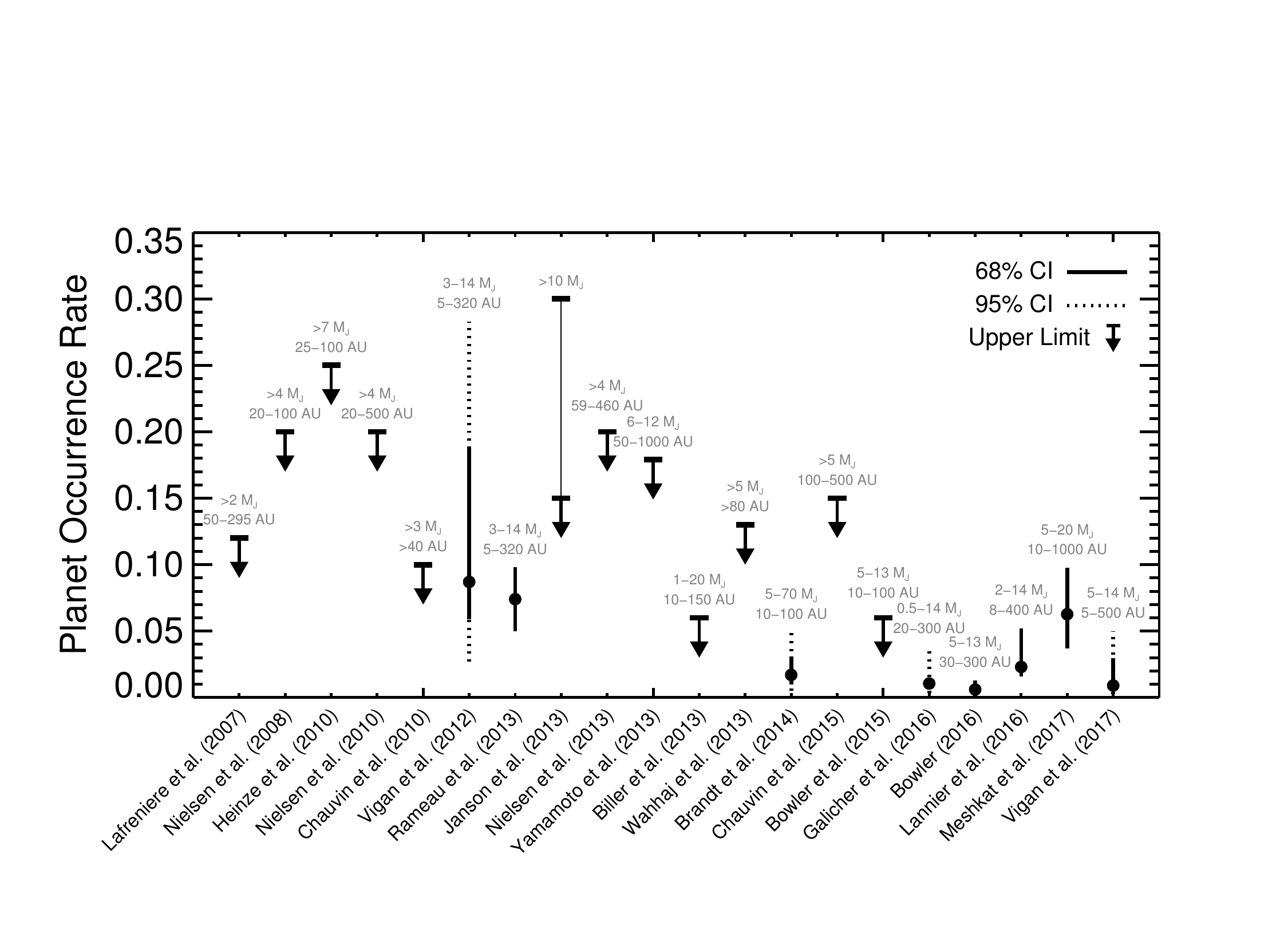}
\vskip -.3in
\caption{Occurrence rate measurements of giant planets from direct imaging surveys over the past decade.
Most surveys have reported upper limits, but larger meta-analyses over the past few years are converging
on a frequency of about 1\% averaged over all host star spectral types.
Arrows indicate upper limits with horizontal bars representing the upper limit value.  
For measurements (as opposed to upper limits), solid and dotted lines denote 68\% and 95\% credible intervals,
respectively.
Note that these surveys are not all independent; targets may overlap and many surveys incorporate previously published results into their statistical analysis. }
\label{fig:plfreq}       
\end{figure}

The frequency of massive planets on wide orbits has progressively come into focus over the past decade
(Figure~\ref{fig:plfreq}; see also \citealt{Bowler:2016jk} for a detailed overview). 
Giant planets at large orbital distances are unquestionably rare, especially in relation 
to lower-mass planets being found with radial velocity and transit methods.  
This dearth of widely-separated gas giants gradually became apparent with the conclusion of  
large-scale surveys carried out at 8--10-meter class facilities over the past 15 years with a wide range of
instruments, most notably by \citet{Biller:2007ht},
\citet{Lafreniere:2007cv}, \citet{Heinze:2010ko}, \citet{Chauvin:2010hm}, 
\citet{Vigan:2012jm}, \citet{Rameau:2013it}, \citet{Yamamoto:2013gu},  \citet{Janson:2013cjb}, 
\citet{Biller:2013fu},  \citet{Nielsen:2013jy},  \citet{Wahhaj:2013iq},  \citet{Brandt:2014hc}, 
\citet{Bowler:2015ja},  \citet{Chauvin:2015jy},  \citet{Galicher:2016hg}, \citet{Lannier:2016eo}, \citet{Uyama:2017dz},
and \citet{Naud:2017fk}.

Merging discoveries and detection limits from published surveys into ever-larger samples has become standard practice 
in order to improve the precision of occurrence rate measurements.
These large meta-analyses  carry the most statistical weight and are necessary to test for potential
correlations with stellar properties and environmental context, which are ultimately expected to 
yield clues about planet formation 
and migration pathways.  The following results are based on compilations of several surveys and 
assume hot-start evolutionary models to infer planet masses and sensitivities.

The first large-scale compilation of direct imaging surveys was carried out by \citet{Nielsen:2008kk}
and expanded to over 118 stars in \citet{Nielsen:2010jt},
who measured an upper limit of $<$20\% on the frequency of $>$4~$M_\mathrm{Jup}$  
planets between $\approx$30--500~AU 
around Sun-like stars (at the 95\% confidence level).
The next major milestone for population statistics was carried out by \citet{Brandt:2014cw}.  They merged
five surveys totaling a sample of 248 unique targets with spectral types from B to mid-M
and found that 1.0--3.1\% of stars host 5--70~$M_\mathrm{Jup}$ 
companions between 10--100~AU (at 68\% confidence; the 95\% credible interval spans 0.52--4.9\%).
In addition, they conclude that the substellar companions from these surveys are consistent with 
a smooth companion mass and separation distribution 
($p(M,a)$ $\propto$ $M^{-0.65\pm0.60}$$a^{-0.85\pm0.39}$), 
implying that the planetary-mass companions identified in these surveys may 
represent the low-mass tail of brown dwarfs rather than a separate
population of high-mass planets.
Recently, \citet{Vigan:2017kb} compiled the largest collection of imaging results 
focusing on FGK stars.  Based on a sample of 199 targets from 12 previous imaging surveys,
they derive a frequency of 0.90\% for 5--14~$M_\mathrm{Jup}$ companions spanning 5--500~AU
(the 68\% credible interval is 0.70--2.95; the 95\% range is 0.25--5.00\%).
For the entire substellar mass range from 5 to 75~$M_\mathrm{Jup}$, they find a somewhat higher
occurrence rate of 2.40\% (with a 95\% credible interval of 0.90--6.80\%).
By quantitatively comparing their results with population synthesis models of planets formed via disk instability,
they lay the groundwork for direct population-level tests of planet formation theory and
conclude that this mode of planet formation is probably inefficient at forming giant planets.
At present there are no signs that host star multiplicity dramatically impacts the 
occurrence rates of circumbinary planets on wide orbits compared to single stars.
\citet{Bonavita:2016kb} measure a frequency of 1.3\% for companions spanning 2--15~$M_\mathrm{Jup}$
between 10--500~AU based on 117 close binary systems (their 95\% confidence range spans 0.35--6.85\%),
which is similar to the rate for their control sample of 205 single stars compiled from the literature.

The largest statistical analyses spanning all host star masses 
were assembled by \citet{Bowler:2016jk} and \citet{Galicher:2016hg}.
Based on results from nine published surveys totaling 384 unique and single young stars, 
\citet{Bowler:2016jk}  found that the overall occurrence rate of 5--13~$M_\mathrm{Jup}$ 
giant planets spanning 30--300~AU is 0.6$^{+0.7}_{-0.5}$\% (68\% credible interval).  For an even wider
range of separations from 10 to 1000 AU, the occurrence rate is 0.8$^{+1.0}_{-0.6}$\%.
This is in good agreement with results from \citet{Galicher:2016hg}, who find
a frequency of 1.05$^{+2.8}_{-0.7}$\% (95\% credible interval)
for 0.5--14~$M_\mathrm{Jup}$ planets between 20--300 AU 
based on their sample of 292 stars spanning B stars to M dwarfs.
Altogether, the most massive giant planets reside around roughly 1\% of stars in the interval
between a few tens to a few hundred AU.

Planetary-mass companions exist at even wider separations with comparably small frequencies.
\citet{Durkan:2016ib} carried out an analysis of $Spitzer$/IRAC observations of young stars 
and found an upper limit of 9\% for 0.5--13~$M_\mathrm{Jup}$ companions between 100--1000 AU.
\citet{Naud:2017fk} measure a frequency of 0.84$^{+6.73}_{-0.66}$\% (at 95\% confidence intervals)
for 5--13~$M_\mathrm{Jup}$ companions between 500--5000 AU, similar to the substellar
companion frequency at ultra-wide separations estimated by \citet{Aller:2013bc}.

With these large samples in hand, the next natural  step is to begin searching for correlations with
stellar parameters and environment to better understand the context in which this 
population of planetary-mass companions forms.
Several surveys have specifically focused on high-mass stars
(\citealt{Ehrenreich:2010dc}; \citealt{Janson:2011hu}; \citealt{Vigan:2012jm}; \citealt{Nielsen:2013jy})
and low-mass stars (\citealt{Delorme:2012bq}; \citealt{Bowler:2015ja}; \citealt{Lannier:2016eo})
to examine the occurrence rate of giant planets as a function of stellar host mass.
\citet{Lannier:2016eo} find intriguing hints of a trend with stellar host mass, but 
this has not been recovered with larger samples by \citet{Bowler:2016jk} and \citet{Galicher:2016hg}.
Breaking their sample into spectral type bins, \citet{Bowler:2016jk} measures an occurrence rate
of 2.8$^{+3.7}_{-2.3}$\% from 110 young BA stars,  $<$4.1\% from 155 FGK stars, 
and $<$3.9\% from 119 M dwarfs at separations of 30--300 AU.  This  is suggestive of a positive correlation
with stellar host mass, but larger samples are still needed to unambiguously 
validate this with statistical rigor.

Following the discovery of the planets orbiting HR 8799 and $\beta$~Pic, several surveys focused 
their attention on stars hosting debris disks in anticipation of a trend with wide-separation planets
(\citealt{Janson:2013cjb}; \citealt{Rameau:2013it}; \citealt{Wahhaj:2013iq}; \citealt{Meshkat:2015dh}).
The best evidence for a correlation with debris disks is presented by \citet{Meshkat:2017jka}.
They assemble a sample of new and published results of 130 single stars hosting debris disks
and compare this with a control sample of 277 stars, ensuring the age and host star mass distributions
are similar.  Meshkat et al. measure a planet frequency of 6.3\% for the debris disk sample with
a 68\% confidence interval spanning 3.7--9.8\%--- which is higher than the control sample with 88\% certitude.
While a firm confirmation is needed with larger samples, this is the best evidence for \emph{any}
correlation with host star properties that has emerged thus far with direct imaging.

\section{Occurrence Rate of Brown Dwarf Companions}

Brown dwarfs are more luminous than giant planets at the same age, so brown dwarf companions to stars have naturally been detected in higher numbers than planets in direct imaging surveys.  Numerous stars with brown dwarf companions have been identified by surveys geared toward finding exoplanets (e.g., PZ Tel B, \citealt{Biller:2010ku}, \citealt{Mugrauer:2010cp}; 1RXS~J235133.3+312720 B, \citealt{Bowler:2012cs}; HD 984 B, \citealt{Meshkat:2015hd};  HR 2562 B, \citealt{Konopacky:2016dk}; HD 206893 B, \citealt{Milli:2017fs}).
Like giant planets, the overall frequency of brown dwarf companions to stars is relatively low (below $\approx5$\%), the luminosity of brown dwarfs decreases with age, and $\sim$1 hour of AO-fed time on large telescopes is usually required for robust PSF subtraction of each star.  As a result, the yield of brown dwarf companions from direct imaging surveys is relatively low compared to stellar companions, but rather high compared to giant planets, making detailed determinations of population properties challenging.

The first direct imaging surveys were limited by the modest inner working angles of the early instruments and coronagraph systems, and therefore first detected the ``low hanging fruit'' of wide-separation brown dwarfs, like GJ 229 B at an angular distance of 7.7$''$ (\citealt{Nakajima:1995bb}) and HR 7329 B at 4.2$''$ (\citealt{Lowrance:2000ic}).  These initial surveys found low occurrence rates for these long period brown dwarfs: \citet{Oppenheimer:2001kl} surveyed 107 stars and detected a single brown dwarf, GJ 229 B, while \citet{Hinz:2002jb} did not discover any substellar companions in their survey of 66 stars with sensitivities to brown dwarfs between 100 and 1400 AU.  A larger sample of 102 stars by \citet{McCarthy:2004hl} was sensitive to planetary-mass companions between 75 and 300 AU, and a further 178 stars where brown dwarf companions between 140 and 1150 AU could be detected.  Only the single candidate substellar companion GL 577 B was discovered but  was later shown to be a low-mass star (\citealt{Mugrauer:2004daa}).  These low detection rates made a robust measurement of the substellar occurrence rate difficult.

Later surveys focused on younger targets and utilized instruments capable of imaging brown dwarfs much closer to the host star (to within $\sim$1$''$), and could thus reach a broader population of brown dwarfs.  With larger samples and greater sensitivity, these studies were better able to directly measure the frequency of brown dwarf companions.  \citet{Metchev:2009ky} surveyed 266 stars, including a deep sample of 100 stars with two detections, resulting in a brown dwarf companion frequency of 3.2$^{+3.1}_{-2.7}$\% between 13--75 M$_\mathrm{Jup}$ and 28--1590 AU.  A similar value was measured by \citet{Brandt:2014cw} based on giant planets and brown dwarfs discovered by the SEEDS program and several previous surveys.  
Similarly, \citet{Vigan:2017kb} find a frequency of 2.45\% for 5--75 $M_\mathrm{Jup}$ companions between 5--500 AU 
(the 95\% credible interval is 0.90--6.95\%).
This occurrence rate appears to be largely independent of age: \citet{Ireland:2011id} estimate a frequency of $\approx$4\% for substellar companions between $\sim$200--500 AU, 
\citet{LaFreniere:2014dj} measure a substellar frequency of 4.0$^{+3.0}_{-1.2}$\% for 5--40 $M_\mathrm{Jup}$
companions between 250--1000 AU,
\citet{Uyama:2017dz} found that two out of 68 stars younger than 10 Myr host brown dwarfs, and 
\citet{Cheetham:2015es} measure a frequency of 7$^{+8}_{-5}$\% for brown dwarfs between 1.3 and 780 AU around $\sim$2 Myr $\rho$ Ophiuchus members.  Interestingly, the analysis by \citet{LaFreniere:2014dj} suggests that
brown dwarf companions are more common at wider separations ($>$250 AU) with 88\% credibility.

Brown dwarf companion frequencies appear to be similar across a broad range of spectral types.  70 B and A stars imaged by the NICI Campaign (\citealt{Nielsen:2013jy}) resulted in three brown dwarfs around these stars, 
while \citet{Bowler:2015ja} observed a similar yield from the PALMS survey of M stars: 4 brown dwarfs out of 78 single stars (2.8$^{+2.4}_{-1.5}$\% for 13--75 $M_\mathrm{Jup}$ companions between 10--100 AU).  \citet{Lannier:2016eo} combined a 58-star VLT/NACO imaging survey with literature results, and found that while there is tentative evidence for a higher frequency of low mass companions around higher mass stars for companions with mass ratios less than 0.01, for intermediate mass ratios of 0.01--0.05 (corresponding to brown dwarf masses between 10 and 52 M$_{Jup}$ for a solar mass primary) there is no evidence for a stellar mass dependence.  Larger sample sizes will be required to robustly determine the extent to which stellar mass impacts the frequency of brown dwarf companions.

With increasing sample sizes and overall number of discoveries, it is possible to move beyond simply measuring occurrence rates of brown dwarfs to also measure the form of the companion mass function.  \citet{Brandt:2014hc} found a single power law fit masses between 5 and 70 $M_\mathrm{Jup}$, based on seven detected substellar companions in their sample.  In contrast, \citet{Reggiani:2016dn} fit a two-population model to results from
the VLT/NACO Large Program and the literature, 
with overlapping planet and stellar companion distributions giving a minimum value in frequency between 10 and 40 $M_\mathrm{Jup}$.  

The in-progress Gemini Planet Imager Exoplanet Survey (GPIES, \citealt{Macintosh:2015ewa}) and SpHere INfrared survey for Exoplanets (SHINE, \citealt{Feldt:2017id}) are both targeting $>$500 stars with next-generation imagers capable of detecting brown dwarfs even closer to their parent stars.  Already these programs have detected new brown dwarfs around HR 2562 (\citealt{Konopacky:2016dk}) and HD 206893 (as part of the SHARDDS campaign, \citealt{Milli:2017fs}); the final surveys will be well-placed to make a more robust measurement of substellar occurrence rate and determine the nature of the stellar mass and companion mass distributions.

\section{Conclusions}

Substantial progress has been made over the past decade constraining the statistical properties of giant planets 
and brown dwarf companions via direct imaging.
Altogether, averaged across
all host star spectral types (B stars to M dwarfs), 
the most precise measurements of the giant planet occurrence rate 
spanning masses of  $\approx$5--13~$M_\mathrm{Jup}$ for the entire range of separations accessible to 
direct imaging ($\approx$5--500 AU) is converging to a value near 1\%.
More nuanced correlations are just beginning to be explored and there are initial indications that
giant planets positively scale with the presence of debris disks, and (to a lesser extent) host star mass.

The frequency of brown dwarf companions (13--75~$M_\mathrm{Jup}$) 
is consistently found to be between $\approx$1--4\%--- 
albeit with considerable uncertainty among individual measurements---
with no obvious signs that this value evolves with age or as a function of stellar host mass.
This indicates that the frequency of brown dwarf companions is comparable to those of
giant planets, with hints that brown dwarfs may exceed planets by a factor of a few.

There remain many open questions that will be especially suitable to examine when the 
large ($>$500 star) surveys with second-generation AO instruments conclude, and 
in the longer-term with the 30-m class of ground-based extremely large telescopes.
These include refining the functional form of the companion mass function and companion mass
ratio distribution; more robust tests for correlations with stellar host mass, the presence of debris disks,
and multiplicity; searching for signs of evolution in the overall occurrence rate or outer separation cutoff
over time; a better understanding of planet multiplicity at wide separations; and 
eventually completing the bridge between the radial velocity and direct imaging planet distribution functions.
The steady improvement of occurrence rate measurements and populations statistics 
is a testament to the communal effort required to build ever-larger samples and 
address increasingly rich questions about planet formation, architectures, and evolution.

\bibliographystyle{spbasicHBexo}  

\end{document}